**Heterointerface-Engineered Electrochemically Exfoliated $MoS_2/WS_2$ 2D-Layered Nanocomposite for Efficient Visible-Light Photocatalytic Degradation of Sorafenib**

I. Agnes Felicia Roy [b,c], Kuo Yuan Hwa*[a,b,c,d,e], Aravindan Santhan[a,b,d], and Slava V Rotkin[f]

[a]Department of Molecular Science and Engineering, National Taipei University of Technology, Taipei, Taiwan (ROC)

[b]Graduate Institute of Organic and Polymeric Materials, National Taipei University of Technology, Taipei, Taiwan (ROC).

[c]Graduate Institute of Energy and Optoelectronic Materials, National Taipei University of Technology, Taipei, Taiwan (ROC).

[d]Center for Technology Impacts and Sustainability, National Taipei University of Technology, Taipei, Taiwan (ROC).

[e]College of Engineering, National Taipei University of Technology, Taipei, Taiwan (ROC).

[f]Materials Research Institute and Department of Engineering Science & Mechanics, The Pennsylvania State University, PA 16802, United States

**Corresponding author**

*Professor. Kuo-Yuan Hwa, Email: kyhwa@mail.ntut.edu.tw

Phone number: 02-27712171 ext.2419 (0), 2439, 2442 (lab).

**Abstract:**

The increasing prevalence of pharmaceutical contaminants within the aquatic environment has generated considerable environmental concerns, especially regarding persistent anticancer medications like the kinase inhibitor sorafenib (SRF), which are inadequately eliminated by standard degradation methods. A heterointerface-engineered $MoS_2/WS_2$, 2D/2D layered nanocomposite was fabricated using an electrochemical exfoliation method to facilitate effective visible-light-driven photocatalytic degradation of SRF. The electrochemical exfoliation method yielded ultrathin 9.62-layer thickness $MoS_2/WS_2$ nanosheets with numerous exposed edge sites and an increased specific surface area, facilitating the development of well-interconnected van der Waals heterointerfaces. Comprehensive structural and morphological examinations utilizing field emission scanning electron microscopy (FE-SEM), atomic force microscopy (AFM), Raman spectroscopy, and UV-visible spectroscopy validated the effective synthesis of few-layer nanosheets and their heterostructure interfaces. In contrast to the pure $MoS_2$ and $WS_2$ nanosheets, the $MoS_2/WS_2$ heterostructure composite demonstrated significantly enhanced photocatalytic efficacy, attaining roughly 92 % degradation of SRF within 2 h under visible-light exposure. The enhanced catalytic efficiency is mainly due to the establishment of a Type-II band alignment at the $MoS_2/WS_2$ interface, facilitating effective charge separation and directional charge transfer while inhibiting electron–hole recombination. The robust interfacial interaction among the transition metal dichalcogenide layers accelerates visible-light absorption and the production of reactive oxygen species. This study illustrates that electrochemically exfoliated $MoS_2/WS_2$ heterostructure composites serve as a viable catalytic platform for the successful removal of persistent pharmaceutical pollutants.

**Graphical Abstract**

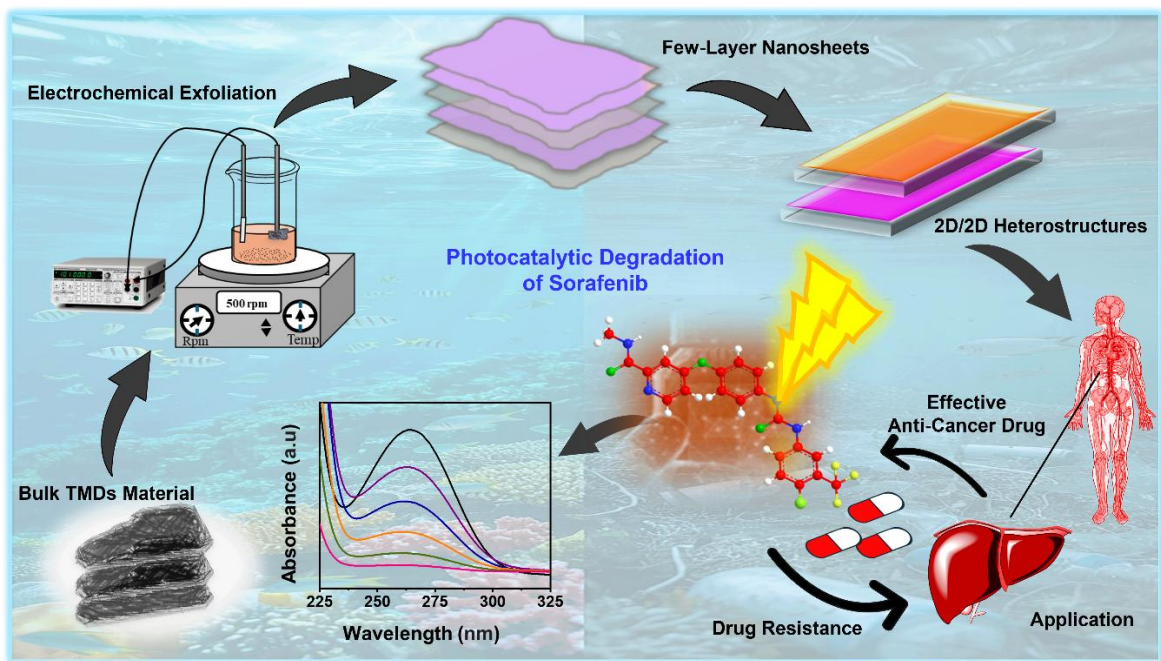

**Introduction**

Pharmaceutical residues are a class of emerging pollutants now recognized for their persistence and biological activity in aquatic ecosystems, and thus pose an ecological and human health risk[1], [2], [3], [4], [5]. As these compounds are generally too complex for conventional treatment systems (such as wastewater treatment plants (WWTPs), many can pass through treatment barriers nearly unaffected, reaching surface and groundwater [1], [3], [6]. Pharmaceutically active compounds are routinely detected in WWTP effluents from nanograms to micrograms per liter by worldwide monitoring programs, and the contamination has been observed in North America, Europe, Oceania, Africa, and Asia[2], [3], [7]. While the main sources of pollutants are effluents from hospitals and pharmaceutical manufacturing plants, domestic wastewater, and runoff from livestock farms, there are numerous ways the pollutants are introduced into the environment, including, but not limited to, directly into water, seepage from soil treated with sludge, and disposal of unwanted medicines[4], [7], [8].

Recent reviews demonstrate substantial regional differences in the composition and concentration of pharmaceuticals, with some parts of Africa having exceptionally high detection frequencies and lacking effective wastewater treatment[3], [7]. Of particular concern is the prevalence of anticancer agents within pharmaceutical contaminants, owing to their intrinsically high toxicity, structural complexity, and resistance to degradation[1], [3]. They function by interfering with cellular functions such as replication and signal transduction, and as such, can pose a significant risk to organisms at extremely low concentrations[9]. Sorafenib (SRF), used in the treatment of hepatocellular and renal cell carcinoma (as a first-line therapy or adjuvant therapy), functions as a multikinase inhibitor and has been used as an example for removal by photocatalysis. [10]

Sorafenib (SRF), is poorly biodegradable and has high chemical stability due to its aromatics and heteroatoms, making its removal from wastewater difficult. While SRF is metabolized, it is excreted partially unmetabolized as well as in active metabolite forms[3], [11]. Its physicochemical properties are important to consider: solubility, polarity, pKa, and soil/water interaction behavior can all impact the persistence and behavior of pharmaceutical residues in the environment; transformation products are often persistent, can become bioactive, or bioaccumulate in non-target organisms[2], [3]. Existing anticancer studies indicate removal of many at WWTPs can be as low as 50% and as high as 90%. One might expect this to be lower for anticancer drugs, given their increased chemical stability and structural complexity[2], [8], [12].

Adverse effects have been observed in aquatic organisms, such as disruptions to the endocrine system, alterations in behavior, problems with reproduction and development in fish, and numerous other phenomena in aquatic invertebrates[2]. Ecological risk assessments for the aforementioned anticancer drugs have demonstrated that high risk quotients can exist, suggesting possible ecological risks at environmentally relevant concentrations[13], [14]. Conversely, antibiotic contamination in the environment contributes to the rise and prevalence of AMR and resistance genes that can move to human pathogens[3], [8]. Cancer drugs, while not directly related to the development of AMR like antibiotics, can have a multitude of adverse effects due to their nature of cell disruption and toxicity, which is particularly worrisome with non-target species and mix-effects when combined with other chemicals[2]. Anticancer agents have been detected in WWTPs and surface waters, with studies reporting accumulation in fish tissue and macroinvertebrates near WWTP discharge sites. For this reason, more specific treatment for anticancer agents like SRF is needed[2], [3], [15].

Advanced oxidation processes, including photocatalysis, are beneficial for the complete mineralization of the persistent contaminants[1], [16], [17]. Photocatalysis is a method of degradation whereby photogenerated electrons and holes formed by a semiconducting material under irradiation react with water and oxygen to form reactive oxygen species (ROS)[18], [19]. The ROS, namely hydroxide and superoxide, are potent oxidizing agents capable of mineralizing complex molecules into carbon dioxide, water, and other inorganic products[16], [17], [20]. Compared to other conventional treatment methods (adsorption, membrane filtration), photocatalysis can completely break down and mineralize recalcitrant organic pollutants. Combining two processes, such as photocatalysis with electrochemical oxidation or Fenton-like reactions, can be synergistic for generating higher amounts of ROS, thus yielding faster reaction times[1], [16]. The efficiency of such systems depends heavily on several variables, including the type of photocatalyst used, light intensity, operating pH, and pollutant concentration, and importantly, integration with existing WWTP infrastructure[21], [22].

Conventional WWTPs exhibit only a modest removal of pharmaceuticals (10-60% of some), and only above 90% of particular pharmaceuticals can be achieved under the most aggressive (advanced) treatments [23]. While membrane bioreactors (MBRs) can demonstrate up to 99% removal efficiency, they are energetically demanding, costly to install and operate, and are susceptible to membrane fouling[4], [24]. Low-energy, nature-based solutions like constructed wetlands and phytoremediation also have limitations due to the persistence of many pharmaceuticals, and in particular, anticancer drugs[25]. Cost-effective, visible-light photocatalytic materials with tunable properties that can efficiently remove pharmaceuticals are needed[16], [26], [27].

In this regard, transition metal dichalcogenides (TMDs) have garnered much interest as visible-light photocatalysts[28], [29]. Unlike traditional metal oxides such as TiO$_2$, which absorb mainly in the UV due to large band gaps, the range of which can vary widely from above 3 eV to less than 1 eV[16], [30], TMDs can utilize most of the visible-light spectrum. MoS$_2$ and WS$_2$ possess band gaps in the range of 1.56 and 1.6 eV[26], [31]. Their unique layered 2-D structure and abundance of defects can also enhance their photocatalytic activity due to abundant surface-active sites and effective charge carrier diffusion at heterointerfaces[32], [33]. However, pristine MoS$_2$ and WS$_2$ do face challenges such as their limited electron-hole separation leading to recombination, poor electrical conductivity, and sometimes poor photocatalytic stability.[16], [17], [29]

Heterojunction construction can be a valid strategy to overcome these limitations. Type II heterojunctions can drive electrons into one material and holes into another, suppressing recombination and promoting the efficiency of electron-hole separation[21], [34], [35]. MoS$_2$ and WS$_2$ heterojunctions are of interest due to similar crystal structure, small lattice mismatch, and appropriate band positions that promote type II band alignment[36], [37], [38], [39]. Previous studies have reported greatly enhanced degradation for dyes (e.g., methylene blue) when using MoS$_2$-based heterojunctions, yet there has been little focus on specific pollutants, such as those used in the treatment of cancer, such as sorafenib.[8], [35], [38], [40]

The current study focuses on preparing few layer MoS$_2$, WS$_2$, and MoS$_2$/WS$_2$ heterojunctions through electrochemically assisted exfoliation and utilizing their visible-light photocatalytic efficiency in degrading sorafenib. Electrochemically assisted exfoliation of TMDs produces large specific surface areas with edge-rich sites favorable to pollutant interaction and high conductivities. These MoS$_2$/WS$_2$ heterojunctions are hypothesized to form a strong interfacial interaction that facilitates photo-excited charge separation, resulting in an increase in photocatalytic activity beyond that of either component alone. Combining a biologically active cancer therapeutic drug with a rationally designed heterojunction of a well-established TMD photocatalyst and an industrially viable exfoliation technique should allow for targeted development and remediation of pharmaceutical pollutants.

## 2. Materials and Methods

### 2.1 Materials and Reagents

All chemicals were purchased from commercial sources and used without further purification. The synthesis was performed using Molybdenum disulfide ($MoS_2$, 99% purity), tungsten disulfide ($WS_2$, 98% purity), which were purchased from 2D Semiconductors as the layered precursors for electrochemical exfoliation into few-layer nanosheets. and sorafenib (a tri-fluorinated chemical, 98% HPLC purity) served as the model pollutant for photocatalytic degradation studies. The solvents and reagents that we used were: poly(vinylpyrrolidone) (PVP, M.w = 40,000), deionized water (18.2 MΩ.cm resistivity) acquired from a Milli-Q purification system, acetonitrile (HPLC grade, 99.9%, Merck), and tetrabutylammonium hexafluorophosphate (BuNPF, 98%, Acros Organics). A 0.25M solution of Tetrabutylammonium ($TBA^+$) electrolyte was dissolved in 20ml of acetonitrile (ACN) in a 100ml beaker. The solution was stirred continuously for about 15 minutes to ensure complete dissolution. The solution was stirred continuously for about 15 min to ensure complete dissolution.

### 2.2 Choice of the Material

For the photocatalytic applications, $MoS_2$ and $WS_2$ were chosen due to their synergistic electronic structures and ability to promote visible light response photocatalysis. Both the TMDs own an appropriate band gap with ranges of 1.59-1.62eV for efficient visible light absorption. When integrated into a 2D/2D heterostructure, the individual layers were stacked through Van der Waals interaction, yielding intimate interfacial contact without disturbing the intrinsic crystal structures of the individual layers. The system therefore achieves favorable Type-II band alignment, favoring the space separation of photogenerated electrons and holes, so that the recombination of charge carriers is efficiently suppressed in the heterostructure system[29], [31]. Although $MoS_2$ has shown superior chemical stability and resistance toward basal plane photo corrosion, $WS_2$ offers relatively higher inherent catalytic activity. The coupling of $MoS_2$ with $WS_2$ thus exhibits synergetic effects toward photocatalysis as well as toward stable photo corrosion behavior. Moreover, exfoliation of 2D nanosheets results in many layers with a large specific surface area and abundant edge sites; thus, more active sites are available for the reactions, resulting in good performance[29].

## 2.3 Electrochemical exfoliations

### 2.3.1. Exfoliation of $MoS_2$

Two electrode setups were used. The bulk $MoS_2$ crystal acts as the working electrode, and the Pt electrode works as the counter electrode. Electrodes are placed about 2 cm apart. Electrodes were connected to a GW Instek DC Power Supply. A constant potential of - 8 V was applied in aqueous $TBA^+$ electrolytes to facilitate the intercalation of solvated ions in the van der Waals gap. The process was carried out for 45 minutes, during which visible exfoliation of the $MoS_2$ layer was observed.

## 2.3.2. Exfoliation of WS$_2$ and Heterostructure Formation

A second exfoliation was carried out for WS$_2$ at specific potentials in the same electrolyte medium as for MoS$_2$. The exfoliated WS$_2$ solution was gradually added to the exfoliated MoS$_2$ solution to allow for close stacking between the materials, which can further create defects on the interface of both materials and increase the catalytic activity. To remove the remaining electrolyte, unexfoliated material, and to increase the surface area, the sample was centrifuged for 10 minutes at 3000 rpm to allow segregation of the few-layer MoS$_2$. After the exfoliated nanosheets were removed, the aqueous dispersion of nanosheets was purified by filtration through a Sartolab RF 150 filter with a pore size of 0.2 μm and a capacity of 150 mL. The obtained solution was filtered and dried at room temperature, and finally stored in a dry vial. An aqueous solution of PVP was then added, and the solution was sonicated for 15 min to obtain a stable suspension.

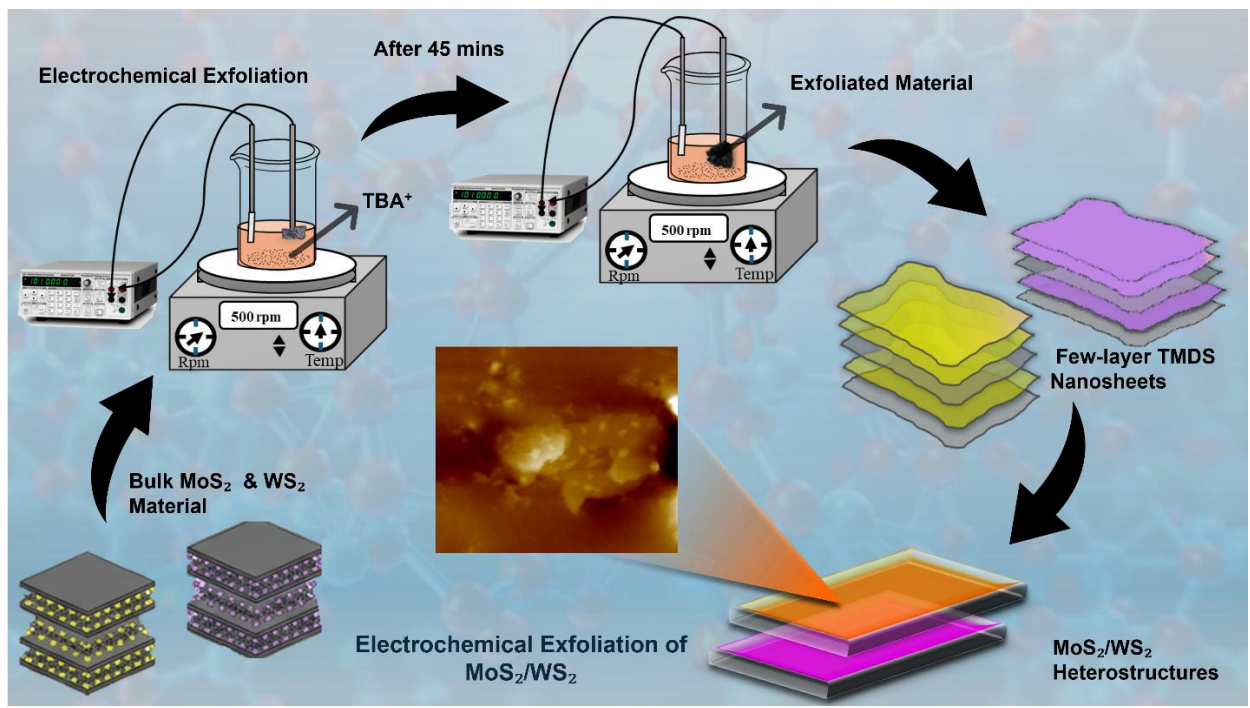

**Scheme 1** depicts the electrochemical exfoliation synthesis scheme of MoS$_2$, WS$_2$, and MoS$_2$/WS$_2$.

## 3. Characterizations and Instrumentations

The morphology and microstructure of electrochemically exfoliated materials, $MoS_2$, $WS_2$, and $MoS_2/WS_2$, were examined using a high-resolution thermal field emission scanning electron microscope (FE-SEM) at an accelerating voltage of 0.1kV ~ 30kV. FESEM provides high-resolution images of the synthesized particles, enabling detailed analysis of their size, shape, and surface structure. To study the thickness of the exfoliated materials, atomic force microscopy was employed. Height profiles extracted from AFM images were used to estimate the thickness and number of layers of the exfoliated $MoS_2$ and $WS_2$ nanosheets and to verify the formation of few-layer structures. The optical properties of the samples were characterized using an Ultraviolet-Vis-NIR Spectrometer. The absorption measurements were performed across the wavelength range of 190~3200 nm with a resolution of ± 0.3 nm (656.1 nm); ± 1.5nm (1312.2nm) to capture the detailed absorption profiles and to evaluate the optical bandgap of $MoS_2$, $WS_2$, and $MoS_2/WS_2$. The crystallinity of the electrochemically exfoliated materials was assessed using a Micro-Raman Spectrometer. The excitation source was a 532 nm laser with a resolution of ~ 1 um. The combination of these characterization techniques has provided a comprehensive evaluation of the structural, morphological, and optical properties of the synthesized $MoS_2$, $WS_2$, and $MoS_2/WS_2$. Photocatalytic degradation experiments were conducted using a batch reactor under UV irradiation. A UV lamp with 0.22 V and 13.1 A was used as the light source. The reaction solution was continuously stirred for 20 minutes in order to maintain a uniform suspension and to ensure maximum contact of the photocatalyst with UV light throughout the process.

## 4. Photocatalytic activity

The photocatalytic activity of MoS$_2$, WS$_2$, and MoS$_2$/WS$_2$ composites was evaluated by a UV lamp with an operating potential of **0.22 V,** and a current of **13.1 A** was used as the light source to irradiate the reaction mixture for monitoring the degradation of SRF, a commonly used anticancer drug. The purpose of this study was to evaluate MoS$_2$, WS$_2$, and MoS$_2$/WS$_2$ composites ability to break down dangerous pharmaceutical contaminants. For Photocatalytic degradation, the experiment was performed by taking 25 mL of sorafenib solution, mixing it with the synthesized photocatalyst, and exposing it to UV irradiation. The absorbance of the solution before irradiation (C$_0$) was recorded with a UV-Vis spectrophotometer. After UV irradiation, aliquots were collected after each 30 min interval and the corresponding absorbance was measured (C$_t$), it was measured to monitor the degradation process. The degradation efficiency (%) was calculated using the equation: [41]

$$Degradation\ (\%) = \left[\frac{C_0 - C_t}{C_t}\right] * 100 \qquad (1)$$

where C$_0$ and C$_t$ represent the initial concentration and the concentration at irradiation time t, respectively. The results obtained in this work could be informative for the practical application of MoS$_2$, WS$_2$, and MoS$_2$/WS$_2$ composites to pharmaceutical pollutant remediation in aquatic environmental systems.

## 4. Results and discussion

## 4.1 Morphological analysis

The Field emission scanning electron microscope (FESEM) of electrochemically exfoliated $MoS_2$, $WS_2$, and $MoS_2/WS_2$ composite nanosheets is shown in **Fig. 1A and B**, which illustrates the morphological properties of the $MoS_2$ samples, which appear thicker with multiple layers on them, which is due to re-stacking during the exfoliation process, leading to the formation of stacked sheets of $MoS_2$.[42]. **Fig. 1C and D** show the exfoliated $WS_2$ samples with a flat surface, thin layered shape with defined edges, and a less aggregated morphology. In **Fig. 1E and F**, it can be clearly observed from the SEM images that, in the $MoS_2/WS_2$ heterostructure, the $WS_2$ ultrathin nanosheets are well decorated and fastened to the underlying $MoS_2$ layered network to construct a 2D/2D heterostructure network. The abundant surface area and effective interfacial contact can be readily generated between $MoS_2$ and $WS_2$, which is favorable for the charge separation and transfer through the heterointerface to improve the visible light photocatalytic activity.

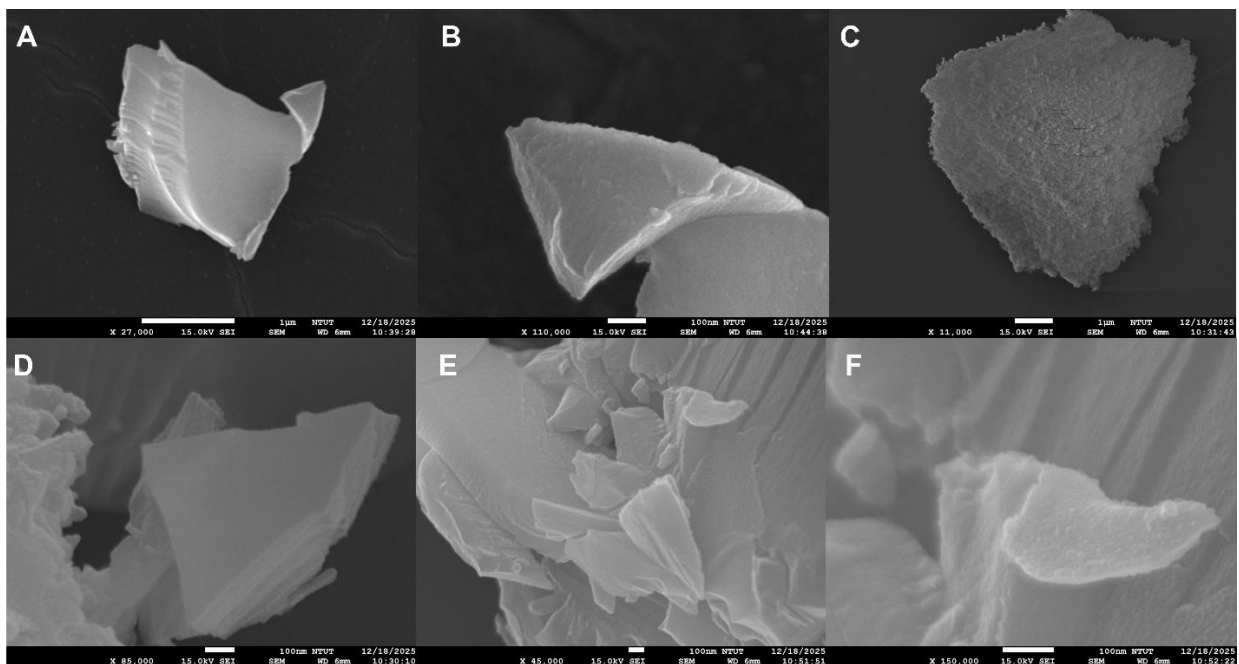

Fig. 1. FESEM images of the exfoliated (A, B) $MoS_2$, (C, D) $WS_2$, (E, F) $MoS_2/WS_2$

**Fig. 2(A–C)** shows the elemental mapping of the MoS$_2$ mix, illustrating the uniform dispersion of Mo and S elements. Specifically, **Fig. 2B** depicts the Mo distribution, and **Fig. 2C** shows the S distribution, confirming the homogeneous composition of MoS$_2$. **Fig. 2(D–F)** presents the elemental mapping of the WS$_2$ mix, revealing the uniform dispersion of W and S elements. **Fig. 2E** displays the W distribution, and **Fig. 2F** shows the S distribution, verifying the stoichiometric integrity of WS$_2$. **Fig. 2(G–J)** illustrates the elemental mapping of the MoS$_2$/WS$_2$ composite mix, confirming the presence of Mo, W, and S [43]. The overall mix elemental mapping in **Fig. 2G** demonstrates proper binding and dispersion of both materials, with individual distributions shown in **Fig. 2H** (Mo), **Fig. 2I** (W), and **Fig. 2J** (S), respectively.

**Atomic Force Microscopy**

AFM was used to investigate the surface topography and thickness of the exfoliated nanosheets shown in **Fig. 3**. A two-Dimensional height map of pure MoS$_2$ flakes over a 1x1 µm scan range showing clean, discrete triangular flakes without much restacking is depicted in **Fig 3A**. In **Fig. 3B**, the 3D surface topography of flakes can be noted, which is ultrathin and has a sharp edge, and **in Fig. 3C,** the height profile of a few-layer structure confirming 10 nm in thickness. Similar homogeneous surface topology is visible in pure WS$_2$ in two-dimensional mode, which is seen in **Fig. 3D**, but high roughness in three dimensions due to the stacking of layers is observable in **Fig. 3E**. The Height profile in **Fig. 3F** depicts a 14 nm thickness, showing significant surface area exposure. In the case of MoS$_2$/WS$_2$ heterostructure, the network of both materials was observed with integrated flakes in **Fig 3G**. **Fig 3H** shows the three-dimensional image of the heterostructures, and it is seen that the overall thickness is 9.62 nm, as seen in **Fig 3I**, which confirms successful van der Waals assembly.[42], [44].

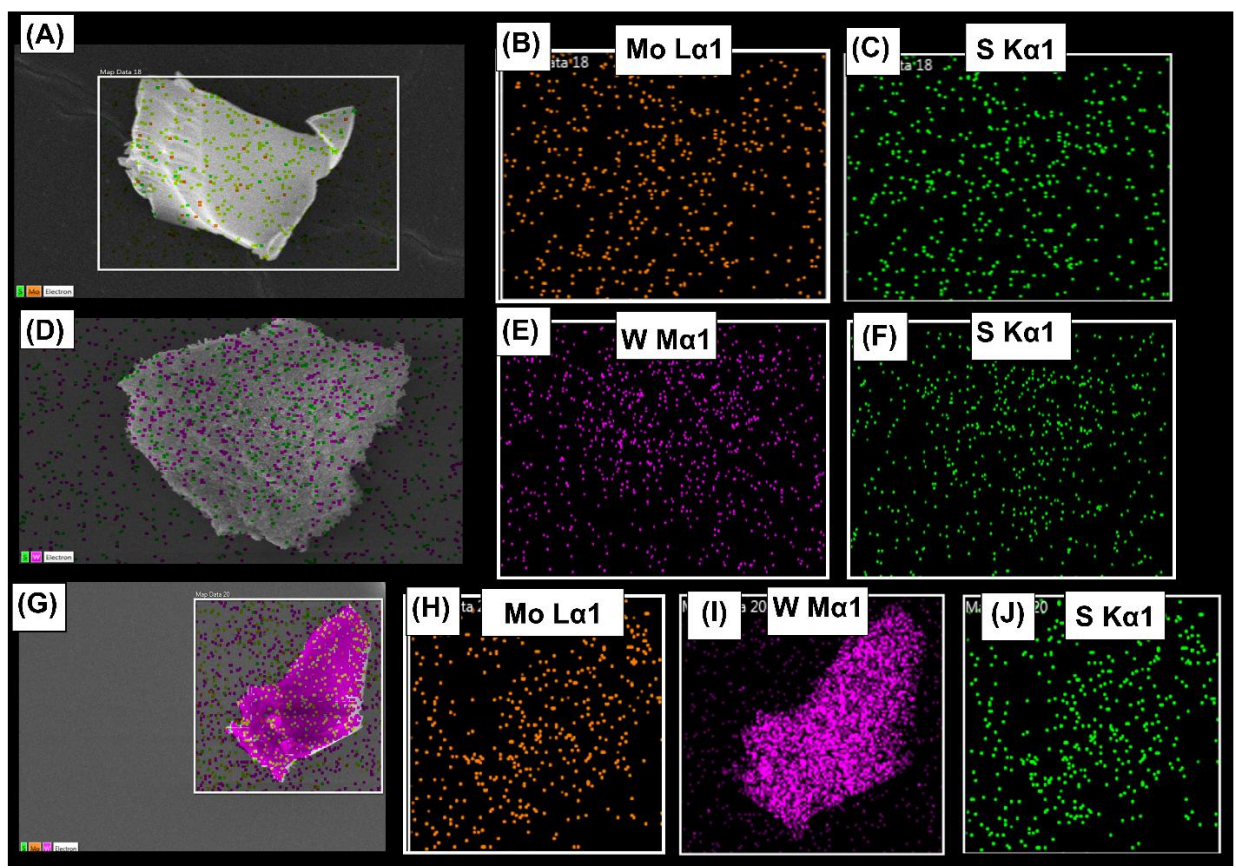

Fig 2. FESEM elemental images of the exfoliated (A-C) MoS$_2$,(C-F)WS$_2$, and (G- J) MoS$_2$/WS$_2$

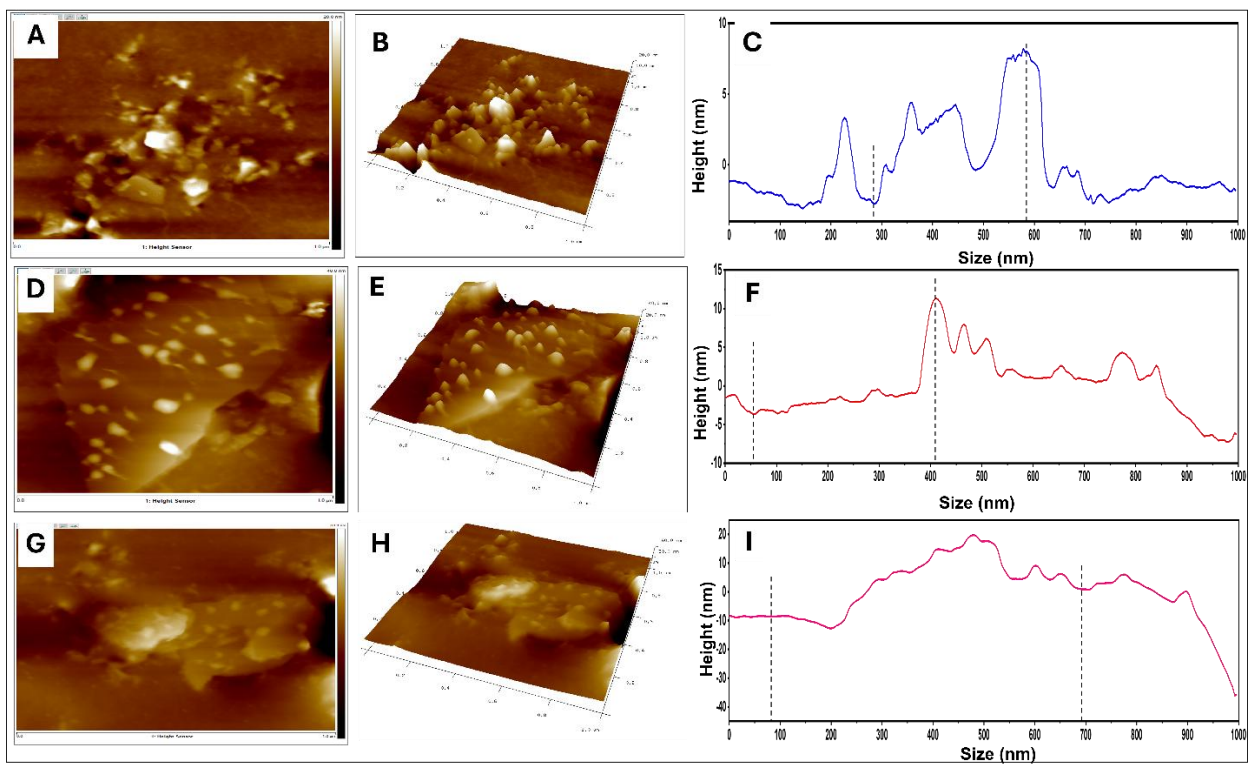

Fig. 3.: Atomic force microscope of electrochemically exfoliated $MoS_2$, $WS_2$, and $MoS_2/WS_2$

## 4.2 Optical Characteristics

### 4.2.1 Raman Spectroscopy

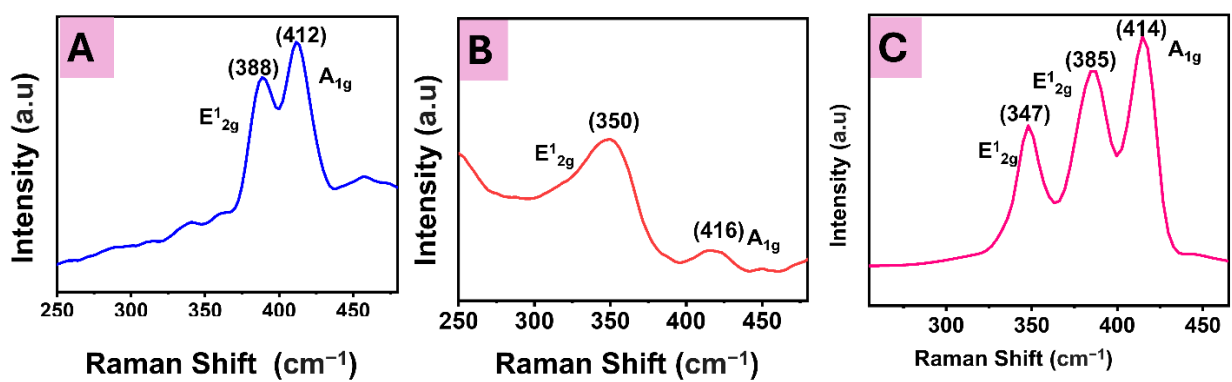

Fig 4: Raman Spectrum of (A) $MoS_2$, (B) $WS_2$, and (C) $MoS_2/WS_2$

Raman spectroscopy was utilized to confirm the structure and layers present in the exfoliated samples[45]. **Fig. 4. (A-C)** displays the spectra for exfoliated MoS$_2$, WS$_2$, and MoS$_2$/WS$_2$, respectively. In **Fig A**, the characteristic peak for MoS$_2$ is located at 388 cm$^{-1}$ and 412 cm$^{-1}$, which correspond to the E$^1_{2g}$ and A$_{1g}$ vibrational modes of MoS$_2$[43]. Similarly, for WS$_2$, the characteristic peaks were observed at 350 cm$^{-1}$ and 416 cm$^{-1}$, which correspond to E$^1_{2g}$ and A$_{1g}$ vibrational modes shown in **Fig B**. In **Fig C,** the MoS$_2$/WS$_2$ composite exhibits well-defined bands associated with the lattice vibrations of both constituents. An intense peak at 347 cm$^{-1}$, which corresponds to the E$^1_{2g}$ vibrational mode of WS$_2$, whereas the peaks around 385 cm$^{-1}$ corresponds to the E$^1_{2g}$ vibrational mode of MoS$_2$ and the prominent peak at 414 cm$^{-1}$ correspond to the overlapping of A$_{1g}$ modes of MoS$_2$ and WS$_2$, which give rise to an intense composite peak in this region and match previous reports of MoS$_2$/WS$_2$ composite van der Waals heterostructure, indicating efficient interlayer coupling at the interface. further confirming the successful formation of the heterostructure.

### 4.2.2 UV-vis spectroscopy:

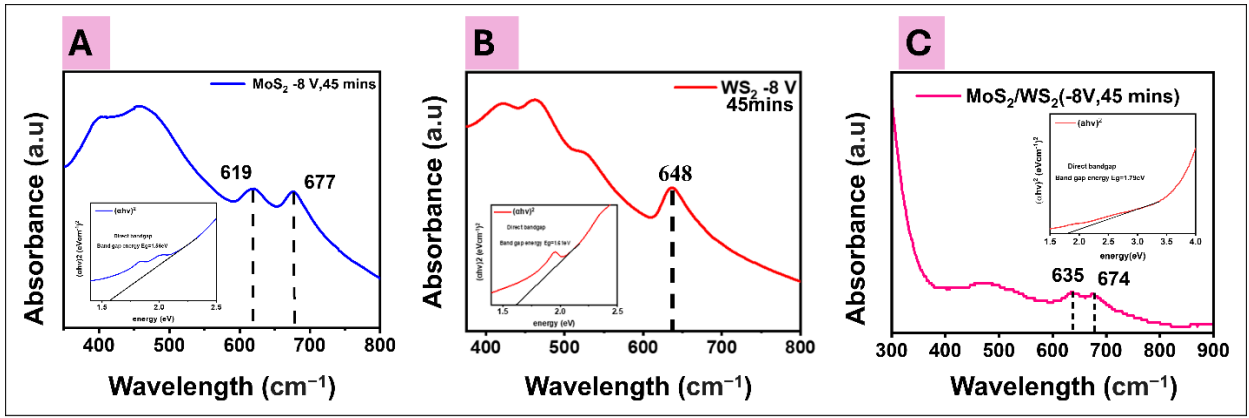

Fig. 5.: The UV-Vis Spectra of (A) MoS$_2$, (B) WS$_2$, and (C) MoS$_2$/WS$_2$

The UV-visible spectroscopy was used to determine the optical absorption and electronic structure of the exfoliated nanosheets, which is shown in **Fig. 5**. $MoS_2$ exhibits two prominent absorption peaks at 619 nm and 677 nm as shown in the **Fig 5 (A),** corresponding to the A and B excitonic transitions of the valence band due to the spin-orbit splitting of the valence band at the K point of the Brillouin zone[46]. These confirm that the original semiconducting 2H phase is well preserved, and few-layered nanosheets have successfully formed by electrochemical exfoliation. Similarly, $WS_2$ had one characteristic excitonic peak at 648 nm as shown in **Fig. 5 (B),** which agrees with its intrinsic electronic structure. From **Fig.5 (C),** it is understood that the $MoS_2/WS_2$ heterostructure has high absorption intensity and a broadened absorption peak across visible light, showing strong electronic coupling at the interface and good light harvesting capability. A blue shift was also seen in the absorption edge in the $MoS_2/WS_2$ heterostructure, which reflects the effect of heterojunction formation on its electronic band structure. The optical band gap (Eg) values were estimated using the Tauc relation:

$$(\alpha h\nu)^2 = A(h\nu - Eg) \qquad [47]$$

where α is the absorption coefficient, hν is the photon energy, A is a constant, and n=2 for direct allowed transitions. The band gap values were determined by extrapolating the linear region of the $(\alpha h\nu)^2$ versus hν plot to the energy axis. The calculated band gaps were 1.64 eV for $MoS_2$, 1.56 eV for $WS_2$, and 1.76 eV for the $MoS_2/WS_2$ heterostructure[42]. The slightly increased band gap of the heterostructure indicates interlayer electronic interaction and band structure modulation induced by heterojunction formation.

## 5. Photocatalytic Degradation of Sorafenib Drug

The anticancer drug sorafenib was used as an example to illustrate the photocatalytic activities of the fabricated $MoS_2$, $WS_2$, and $MoS_2/WS_2$ heterostructure photocatalysts. The photocatalytic degradation experiments were performed in a batch reactor under UV illumination from a lamp operating at 0.22 V and 13.1 A, with constant stirring to prepare a uniform suspension of catalyst and drug. Before irradiation, the mixture of catalyst and SRF was magnetically stirred in the dark for about 20 Minutes until the adsorption-desorption equilibrium was reached and any non-photocatalytic degradation effect was excluded. The value of sorafenib concentration did not change during the dark adsorption-desorption equilibrium, so that the subsequent degradation of the drug would not arise from non-photocatalytic action during UV light irradiation. The degradation of SRF was observed using a UV-Vis spectrometer. Aliquot (1 mL) was taken from the reaction mixture every 30 minutes for analysis until the total irradiation time was 120 minutes, and the SRF absorption band was measured with a UV-Vis spectrophotometer. The decrease in absorbance was used to calculate the residual concentration and degradation efficiency, enabling a quantitative assessment of the photocatalytic activity of $MoS_2$, $WS_2$, and the $MoS_2/WS_2$ heterostructure toward sorafenib.

## 5.4 Comparison Table

| Catalyst | Photocatalyst | Light Source | Time Duration | Efficiency (%) | Reference |
|---|---|---|---|---|---|
| Dabrafenib. | $TiO_2$-based catalyst | UV degradation | 3 days | 99.37 | [48] |
| 5-Fluorouracil | $g-C_3N_4$ composite | Visible light | 100 min | 97.4 | [49] |
| Doxorubicin | ZnO heterostructure | Visible light | 120 min | 93.8 | [50] |
| Sorafenib | $MoS_2/WS_2$ | UV Degradation | 120 mins | 92% | This Work |

Table 3: Degradation efficiency (%) of different Anti-cancer Drug

## 5.1 UV-Vis Absorption and Calibration Study of Sorafenib Drug

A stock solution of SRF (25 ppm) was prepared by accurately dissolving a measured amount of the drug in DW, and ethanol (9:1) was gradually added until the SRF drug became fully soluble, ensuring complete dissolution during preparation. The mixture was stirred continuously using a magnetic stirrer until the drug was completely dissolved, ensuring homogeneity of the stock solution. About 20 mL of the SRF drug was taken, and the absorbance was measured using a UV-visible spectrophotometer at a wavelength of 264 nm. The visible light photocatalytic performance was examined in the degradation of SRF. A (1:1) ratio of the drug and the catalyst was taken, and the absorption was measured. In which the Degradation results were determined using UV-vis spectroscopy by monitoring a continuous decrease in absorbance intensity at 264 nm, which is the wavelength of SRF maximal absorption. The decrease in peak intensity with increasing irradiation time shows that the chromophoric nature of SRF was broken down. The efficiency of degradation was calculated using equation 1[13].

At first, the pure MoS$_2$ and WS$_2$ showed moderate activity under visible-light irradiation. After 120 minutes, the samples got degraded to 62 % and 68 %, respectively, as shown in Table 2, which is due to the recombination of photogenerated electrons and holes. In contrast, **Fig 7 (A)** shows the MoS$_2$/WS$_2$ composite displays better degradation efficiency of about 92.1 %, which benefits from forming a heterojunction. Because of Good visible-light absorption due to narrow band gaps of both MoS$_2$ (1.56 eV) and WS$_2$ (1.64 eV), and also interfacial charge separation. The intimate contact between the layered structure of both materials will benefit the directional transfer of charge carriers, to reduce where C$_0$ and C$_t$ are SRF concentrations at t = 0 and t, respectively. At first, the pure MoS$_2$ and WS$_2$ shore combination. The degradation kinetics obeyed the pseudo-first-order kinetic model, which is represented as follows: Degradation followed pseudo-first-order kinetics:

$$\ln\left(\frac{C_0}{C_t}\right) = kt \qquad [51]$$

Where k is the apparent rate constant (min$^{-1}$) [52]. The Linear plots of ln(C$_0$/C$_t$) vs. time. As shown in the **Fig 7(C),** MoS$_2$ and WS$_2$ nanosheets exhibit moderate photocatalytic activity with the K values of 0.008 min$^{-1}$ and 0.0074 min$^{-1}$, along with R$^2$=0.994 for MoS$_2$ and R$^2$=0.969 for WS$_2$, respectively. In contrast, the MoS$_2$/WS$_2$ heterojunction composite achieved a significantly higher rate constant of 0.01795 min$^{-1}$ (R$^2$ = 0.986), indicating accelerated degradation kinetics. The higher k for the composite is attributed to efficient visible-light absorption, type-II band alignment facilitating charge separation, and enhanced generation of reactive oxygen species (•OH and •$O_2^-$). These results confirm that heterojunction formation substantially improves the photocatalytic performance of the composite compared to individual MoS$_2$ and WS$_2$ nanosheets.

### 5.2 Radical Scavenger (Inhibitor)

The process that explores the ROS that leads the photoreaction of sorafenib, different **reactive oxygen species (ROS) scavengers** were introduced into exfoliated MoS$_2$/WS$_2$ nanosheets, as shown in **(Fig. 7F)**. In this work, **three inhibitors were used** to determine the role of reactive species in the photocatalytic process. Isopropanol (IPA) was employed as a scavenger for hydroxyl radicals (•OH), benzoquinone (BQ) was used to quench superoxide radicals (•$O_2^-$), and ethylenediaminetetraacetic acid (EDTA) was used as a hole ($h^+$) scavenger. We can see the inhibition is very small when adding isopropanol (IPL, •OH scavenger) but a high reduction is found when adding ETDTA-2Na ($h^+$ scavenger) and p-benzoquinone (p-BZQ, •$O_2^-$ scavenger) (decrease about 45 % and 72 %), suggesting that holes ($h^+$) and superoxide radicals (•$O_2^-$) were the major reactive species. This phenomenon is also consistent with absorption shifts and degradation kinetics in **Fig. 7A-D**. Defect-induced traps played an important role in charge separation and promoted ROS generation under visible light.

## 5.3 Stability and reusability of Sorafenib

| S. No | Catalyst | Degradation efficiency % (120 mins, 25 ppm of Sorafenib) | K (min$^{-1}$) | R$^2$ |
|---|---|---|---|---|
| 1 | MoS$_2$ | 62% | 0.008 min$^{-1}$ | 0.994 |
| 2 | WS$_2$ | 68% | 0.0074 min$^{-1}$ | 0.969 |
| 3 | MoS$_2$/WS$_2$ | 92% | 0.01795 min$^{-1}$ | 0.986 |

Table 2: Degradation Efficiency (%)

The recyclability of the photocatalyst is very important for its application, considering economic aspects in combination with stable activity and structure.[53]. The used catalyst was washed in ethanol for the removal of the adsorbed SRF and intermediates, followed by drying at room temperature overnight, and air-drying after each degradation reaction. As shown in **Fig. 7E,** the durability of the composite was very good; it had 92 % degradation of the SRF under optimal conditions in the fifth cycle of reuse. The small activity drop is due to the increasing accumulation of recalcitrant intermediates and unreacted SRF on the catalyst surface, and it blocks the active sites, which is quite typical for photocatalytic removal of pollutants.

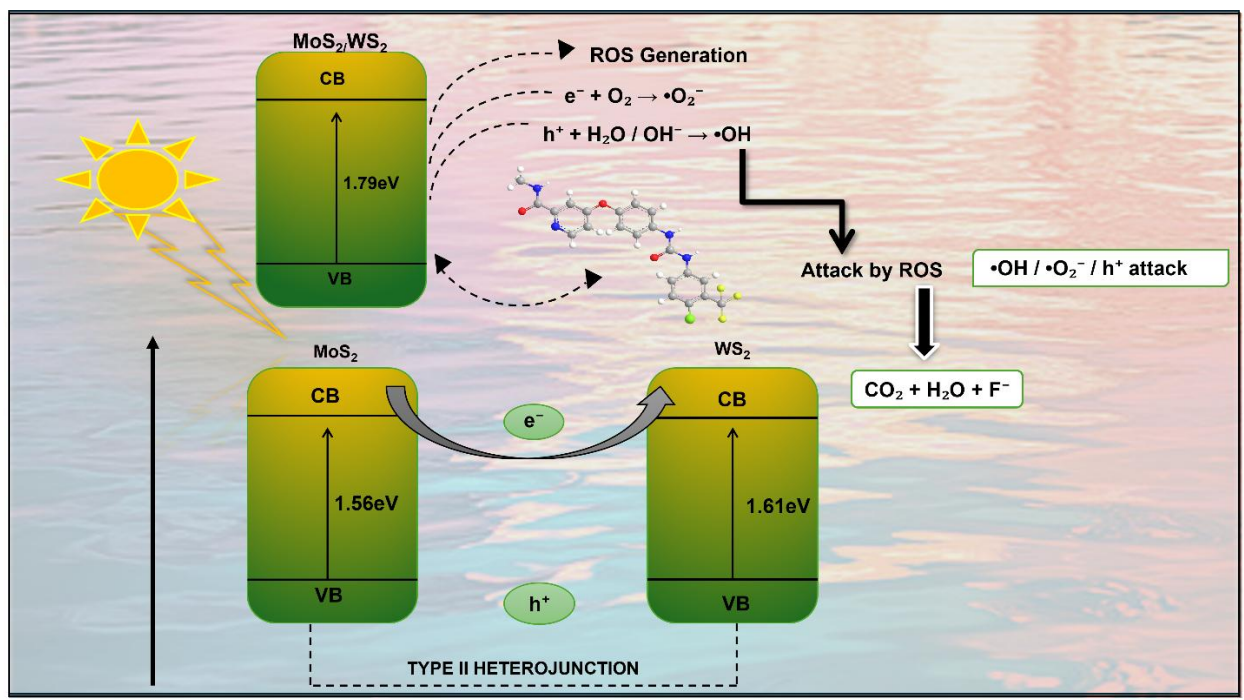

Fig 6: Proposed mechanism for photocatalytic degradation of Sorafenib (SRF) drug

### 5.3 Proposed mechanism of photocatalytic degradation of Sorafenib drug

The Photocatalytic Degradation of SRF using $MoS_2$ and $WS_2$ nanosheets is primarily driven by the generation of reactive oxygen species upon [31]visible light exposure. During the reaction under visible light, photogenerated charges ($e^-$/$h^+$) were created from the electron transition from the valence band (VB) to the conduction band (CB)[54], [55]. Due to the type-II junction of the $MoS_2$/$WS_2$ heterojunction, electrons are transferred from the CB of $MoS_2$ to the CB of $WS_2$, while holes are transferred from the VB of $WS_2$ to the VB of $MoS_2$, thereby increasing the separation of the charge carriers **(Fig 6).** Driven by the photogenerated charges, the following redox reactions took place:

$$e^- + O_2 \rightarrow \bullet O_2^- \qquad [56]$$

$$h^+ + H_2O \rightarrow \bullet OH + H^+ \qquad [57]$$

$$Sorafenib + (\bullet O_2^- + \bullet OH) \rightarrow CO_2 + H_2O \qquad [58]$$

Therefore, the separation of the photogenerated electrons and holes at the interface was achieved. The photoinduced electrons react with dissolved $O_2$ to generate $\bullet O_2^-$ which oxidizes the holes at the surface to OH by the interaction with $H_2O$.[55], [59]. These radical species then degrade SRF into $CO_2$ and $H_2O$. A high generation rate and poor radical recombination lead to excellent photocatalytic performance of the composite catalyst. The generated reactive oxygen species (ROS) effectively oxidize SRF into mineralization products.[60] Comparing the homogeneous and heterostructure 2D materials helps explain their photocatalytic behavior very well. $MoS_2$ and $WS_2$ are single-phase transition metal dichalcogenides, while the heterostructure composite $MoS_2/WS_2$ is comprised of two distinct layered semiconductors. While both $MoS_2$ and $WS_2$ are capable of absorbing visible light and catalyzing photoreactions, their quantum efficiency is limited due to rapid electron–hole recombination. On the other hand, the heterostructure $MoS_2/WS_2$ displays higher activity because the heterojunction interface between the two different layered semiconductors allows for better separation and transport of charge, reduces recombination, and enhances ROS production. Thus, the heterostructure $MoS_2/WS_2$ nanocomposite has superior photocatalytic activity to the homogeneous materials.[31]

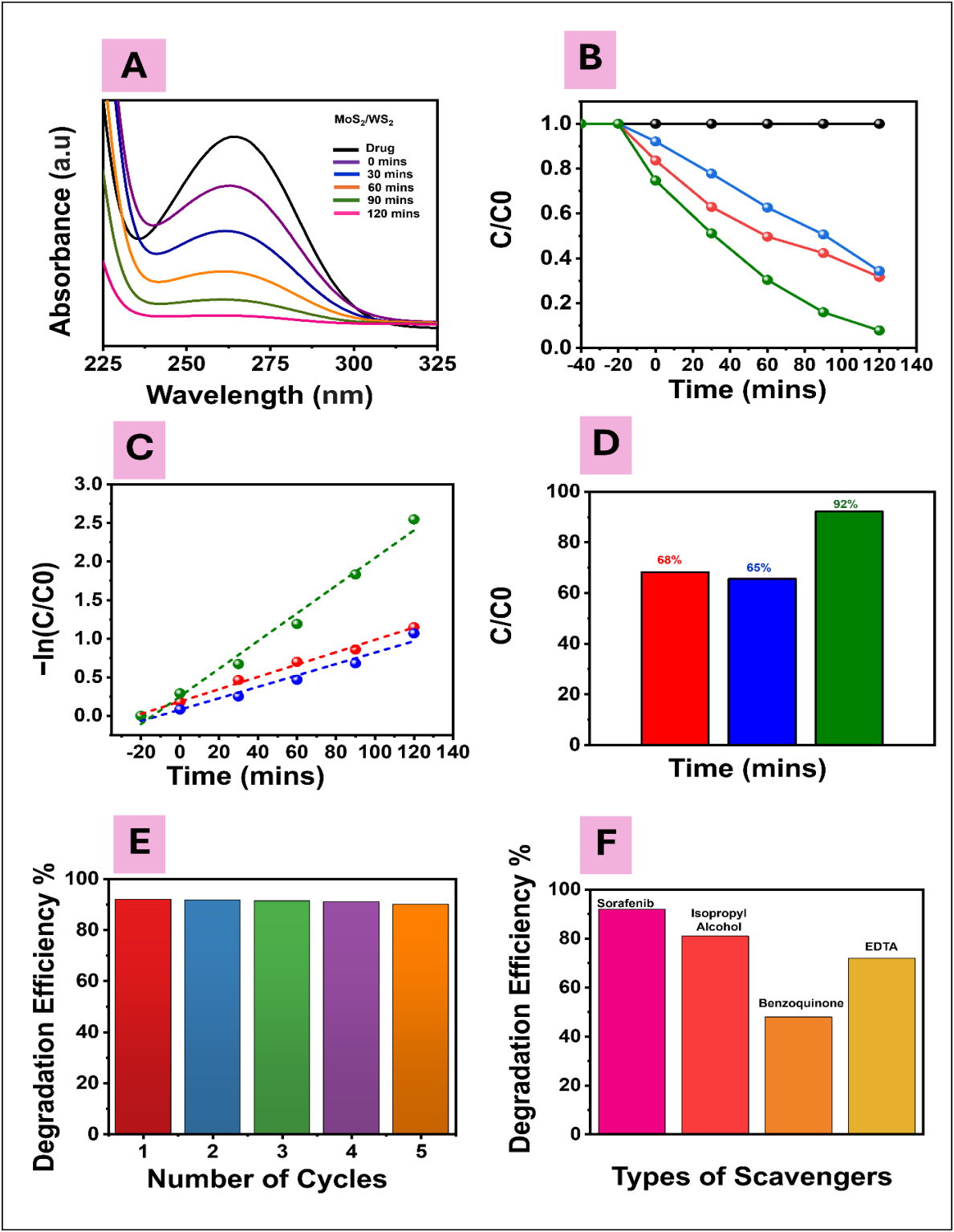

Fig 7: (A-D) photocatalytic activity

**6. Conclusion:**

In Summary, Few-layer $MoS_2$, $WS_2$, and $MoS_2/WS_2$ heterostructure nanosheets were synthesized by electrochemical exfoliation. The morphologies and structure of the nanosheets were characterized by FESEM, AFM, Raman, and UV-Vis, confirming their high crystallinity, layered structure, and strong absorption of visible-light. The enhanced photocatalytic performance in the degradation of SRF of $MoS_2/WS_2$ heterostructure was observed, with nearly 92 % removal in 120 min, and its kinetic rate constant reached 0.01795 min$^{-1}$, nearly three times as large as that of pristine $MoS_2$ and $WS_2$. The good performance is believed to be associated with the Type II heterojunction, leading to efficient charge separation and wide absorption of visible-light and numerous active sites. It is concluded that $MoS_2/WS_2$ heterostructures show high activity, good stability, and visible light activity. This catalyst also exhibited excellent stability and reusability across multiple cycles. These attributes position electrochemically exfoliated $MoS_2/WS_2$ heterostructures as scalable, promising photocatalysts for eliminating persistent pharmaceutical contaminants like SRF, which is primarily manufactured in India and China, from aquatic environments, advancing sustainable wastewater remediation.

**Competing financial interests**: The authors declare no competing financial interests.

**Acknowledgments**

The project gets support from the National Taipei University of Technology. The authors would like to express their gratitude to the Center for Technology Impacts and Sustainability at National Taipei University of Technology for its financial support. The Precision Analysis and Materials Research Center of the National Taipei University of Technology supports the instrumentation.

[48] P. Grover, M. Bhardwaj, L. Mehta, T. Naved, and V. Handa, "Development and Validation of Novel and Highly Sensitive Stability-Indicating Reverse Phase UPLC Method for Quantification of Dabrafenib and its ten Degradation Products," *Indian Journal of Pharmaceutical Education and Research*, vol. 56, no. 3, pp. 888–898, Jul. 2022, doi: 10.5530/ijper. 56.3.142.

[49] I. Rabani, N. T., Tran, M. F., Maqsood, M., Kaseem, G. Dastgeer, and H. B. Truong, "Visible light-driven photocatalytic degradation of rhodamine B and 5-fluorouracil using ZIF-8/GO: unveiling mechanisms," *RSC Adv.*, vol. 15, no. 37, pp. 30217–30230, Aug. 2025, doi: 10.1039/d5ra05177k.

[50] L. R. Rad, M. Irani, and M. Anbia, "Electrocoagulation-photocatalysis sequential combined method for the removal of anticancer drugs from pharmaceutical wastewater using NH2-MIL-125(Ti)/cobalt ferrite nanorods composite photocatalyst," *J. Environ. Chem. Eng.*, vol. 12, no. 5, oct. 2024, doi: 10.1016/j.jece.2024.113302.

[51] H. D. Tran, D. Q. Nguyen, P. T. Do, and U. N. P. Tran, "Kinetics of photocatalytic degradation of organic compounds: a mini-review and new approach," Jun. 05, 2023, *Royal Society of Chemistry*. doi: 10.1039/d3ra01970e.

[52] H. D. Tran, D. Q. Nguyen, P. T. Do, and U. N. P. Tran, "Kinetics of photocatalytic degradation of organic compounds: a mini-review and new approach," Jun. 05, 2023, *Royal Society of Chemistry*. doi: 10.1039/d3ra01970e.

[53] W. S. Koe, J. W. Lee, W. C. Chong, Y. L. Pang, and L. C. Sim, "An overview of photocatalytic degradation: photocatalysts, mechanisms, and development of
37